\documentclass[12pt,a4paper]{article}

\usepackage[dvips]{graphicx}
\usepackage{amsmath,amssymb,amsthm}
\usepackage{bbm,mathbbol}

\newcommand{\Cx}{{\mathbb C}}

\newcommand{\Nl}{\mathbb{N}}

\newcommand{\Rl}{{\mathbb R}}

\newcommand{\idty}{\Eins}

\DeclareMathOperator{\ad}{Ad}

\DeclareMathOperator{\spa}{span}
\DeclareMathOperator*{\tr}{Tr}
\newcommand{\<}{\langle}
\renewcommand{\>}{\rangle}
\providecommand{\abs}[1]{\lvert#1\rvert}
\providecommand{\norm}[1]{\lVert#1\rVert}

\newcommand{\bi}[1]{\boldsymbol{#1}}
\renewcommand{\c}[1]{\mathcal{#1}}
\newcommand{\g}[1]{\mathfrak{#1}}

\renewcommand{\r}[1]{\mathrm{#1}}

\newtheorem{proposition}{Proposition}
\newtheorem{lemma}{Lemma}

\title{A three state invariant}
\author{
M.~Fannes\footnote{E-mail:
mark.fannes@fys.kuleuven.ac.be}, 
D.~Vanpeteghem\footnote{E-mail:
dimitri.vanpeteghem@fys.kuleuven.ac.be}
\footnote{Research Assistant of the Fund for Scientific Research, Flanders
(Belgium) (F.W.O., Vlaanderen)}}

\begin{document}

\maketitle

\begin{center}
Instituut voor Theoretische Fysica \\
K.U.Leuven \\
Celestijnenlaan~200D, B-3001 Heverlee, Belgium 
\end{center}

\begin{abstract}
For triples of probability measures, pure quantum states and mixed quantum
states we obtain the exact constraints on the fidelities of pairs in the
sequence. We show that it is impossible to decide between a quantum model,
either pure or mixed, and a classical model on the basis of the fidelities
alone. Next, we introduce a quantum three state invariant called phase and
show that any sequence of pure quantum states is uniquely reconstructible
given the fidelities and phases.
\noindent 
\end{abstract}

\section{Preliminary}

The transition probability or overlap between states is one of the
fundamental ingredients of quantum mechanics. It lies at the basis of its
probabilistic interpretation. Two states with an overlap close to one are
almost equal. This motivates the term fidelity which will be used as a
synonym of transition probability.

Also in the context of classical probability there is a quite common notion
of distance between probability measures, the Hellinger distance based
on overlaps of densities. This distance is given by $\r d^2_{\r
H}(\mu;\lambda) = 1 - A(\mu;\lambda)$, where
$A$ is called the affinity between $\lambda$ and $\mu$. Affinity
is the classical counterpart of transition probability.

We shall be concerned here with the notion of fidelity in three different
settings: probability measures, pure quantum states and mixed quantum
states. As there are natural inclusions between these three sets extending
the affinity between classical measures to fidelity for mixed quantum
states, we gradually widen our scope when passing from probability measures
to mixed quantum states.

Notions of closedness of states belonging to a sequence generated by a
dynamics or by a coding procedure are obviously important to decide either
on the regular or the chaotic nature of the sequence~\cite{fanspin} or on
compressibility issues, see~\cite{sch,petmos}. In previous work, we used the
spectrum of the Gram matrix of the sequence for this purpose,
see~\cite{decfanspin,fanspin2}. This spectrum is however a complicated function
depending on the full sequence.

The point of this paper is twofold. In Section~\ref{s2}, we analyse
sequences of three states. This is the simplest non-trivial situation. We
obtain the precise constraints on the fidelities of pairs of states in the
sequence for measures and for pure and mixed quantum states. They turn out to
coincide. In other words, on the basis of an admissible triple of fidelities
it is impossible to decide whether one deals with probability measures or
with pure or mixed quantum states.

Next, we introduce in Section~\ref{s3} a quantum three state invariant
called phase and we show that, for pure quantum states, any arbitrary
sequence can be uniquely reconstructed on the basis of the fidelities of
pairs and phases of triples in the sequence.

We recall the basic definitions.

The fidelity between two probability measures $\bi\lambda = (\lambda_1,
\lambda_2, \ldots,\lambda_N)$ and $\bi\mu = (\mu_1,\mu_2, \ldots,\mu_N)$
on an event space with $N$ elements is given by
\begin{equation}
 F(\bi\lambda \,; \bi\mu) := \Bigl( \sum_{j=1}^N \sqrt{\lambda_j\mu_j}
 \Bigr)^2.
\label{1}
\end{equation}
We may extend fidelity to general probability measures by using the Radon-Nikodym
derivative.

For quantum systems, we call expectation functionals on the observables
states. E.g., in standard quantum mechanics a normalised vector $\varphi$ in
the Hilbert space $\g H$ of the system yields a pure state
\begin{equation*}
 P_\varphi(X) := \< \varphi, X\, \varphi\>, \quad \text{$X$ an Hermitian
 operator on $\g H$}.
\end{equation*}   
Writing 
\begin{equation*}
 P_\varphi(X) = \tr \bigl( |\varphi\>\<\varphi|\, X \bigr)
\end{equation*}
we see that $P_\varphi$ is uniquely determined by the one-dimensional
projector on the subspace $\Cx\varphi$ and we shall therefore identify
$|\varphi\>\<\varphi|$ with $P_\varphi$. A (normal) mixed state on $\g H$ is
then a general density matrix $\rho$. 

The fidelity between $P_\varphi$ and $P_\psi$ is the usual quantum mechanical 
transition probability between $\varphi$ and $\psi$
\begin{equation}
 F(P_\varphi \,; P_\psi) := \abs{ \<\varphi,\psi\> }^2 = \tr \bigl( P_\varphi\, P_\psi
 \bigr). 
\label{2}
\end{equation}
For mixed states, Uhlmann extended the notion by using the purification
procedure, i.e.\ by obtaining a mixed states as a marginal of a pure state on
a composite system, see~\cite{uhl}. This leads to
\begin{equation}
 F(\rho \,; \sigma) := \Bigl( \tr \sqrt{\rho^{\frac{1}{2}}\, \sigma\,
 \rho^{\frac{1}{2}}}\Bigr)^2 = \Bigl( \tr \sqrt{\sigma^{\frac{1}{2}}\, \rho\,
 \sigma^{\frac{1}{2}}}\Bigr)^2.
\label{3}
\end{equation}
An isometric transformation $U$ induces the transformation $\rho \mapsto
U\,\rho\,U^* = \rho \circ \ad(U^*)$ on density matrices. The spectrum of $\rho$, taking 
degeneracies into account, remains invariant up to multiplicities of zero.
This is also the case for the fidelity
\begin{equation*}
 F(\rho \,; \sigma) = F(U\,\rho\,U^* \,; U\,\sigma\,U^*) =
 F(\rho\circ\ad(U^*) \,; \sigma\circ\ad(U^*)).
\end{equation*}
Again, fidelity between states can be extended to general quantum
probability spaces, see~\cite{ararag,uhl}. It is straightforward to check that
the three definitions~(\ref{1})--(\ref{3}) of fidelity are compatible.

A simple example is given by states on the matrices of
dimension 2. In this case there is an affine isomorphism between the density
matrices of dimension 2 and the unit ball in $\Rl^3$ explicitly given by the
Bloch transformation
\begin{equation*}
 \rho_{\bi x} := \frac{1}{2} \Bigl(\idty + \bi x\cdot \bi\sigma \Bigr),
 \quad \bi x\in\Rl^3,\ \norm{\bi x}\le 1.
\end{equation*}
Here $\bi\sigma$ are the usual Pauli matrices.
The two dimensional case is rather misleading as the topological and convex
boundaries of the state space coincide, they are namely the unit sphere in
$\Rl^3$. For the $d$-dimensional matrices, the state space has $d^2-1$
real dimensions, its topological boundary $d^2-2$ while the pure state space
has dimension $2(d-1)$. So, the topological boundary has many flat
pieces. The Uhlmann fidelity can be readily computed for the two dimensional
case and one obtains
\begin{equation*}
 F(\rho_{\bi x} \,; \rho_{\bi y}) = \frac{1}{2} \Bigl( 1 + \bi x\cdot\bi y +
 \sqrt{1-\norm{\bi x}^2} \sqrt{1-\norm{\bi y}^2} \Bigr).
\end{equation*}  
In particular, for pure states, i.e.\ $\norm{\bi x} = \norm{\bi y} = 1$, the
fidelity is the square of the cosine of half the angular distance between $\bi x$
and $\bi y$.

\section{Fidelities of triples}
\label{s2}

In this section, we consider a triple of states or measures and compute the
fidelities pairwise. We then look for the constraints between the three
fidelities $F_{12}$, $F_{13}$ and $F_{23}$. Plotting $(F_{12}^{\frac{1}{2}},
F_{13}^{\frac{1}{2}}, F_{23}^{\frac{1}{2}})$ in $\Rl^3$ we obtain subsets
of $(\Rl^+)^3$ when the states run either through the measures, the pure or
the quantum mixed states. These sets are obviously ordered by inclusion.
Moreover they are closed, convex and compact. The convexity follows from the
direct sum construction. E.g.\ for the Uhlmann fidelity 
\begin{align*}
 &\tr \sqrt{\Bigl({\textstyle \frac{1}{2}}(\rho_1 \oplus \rho_2) \Bigr)^{\frac{1}{2}} 
 {\textstyle \frac{1}{2}}(\sigma_1 \oplus \sigma_2) 
 \Bigl({\textstyle \frac{1}{2}}(\rho_1 \oplus \rho_2) \Bigr)^{\frac{1}{2}}}
 \\
 &\qquad\qquad= \frac{1}{2} \Bigl( \tr \sqrt{\rho_1^{\frac{1}{2}}\, \sigma_1\,
 \rho_1^{\frac{1}{2}}} + \tr \sqrt{\rho_2^{\frac{1}{2}}\, \sigma_2\,
 \rho_2^{\frac{1}{2}}} \Bigr).
\end{align*}

\begin{lemma}
\label{lem1}
 Let $\varphi_1$, $\varphi_2$ and $\varphi_3$ be three normalised vectors in
 a Hilbert space $\g H$ and let $F_{jk} := F(P_{\varphi_j} \,;
 P_{\varphi_k})$, then
 \begin{equation}
  F_{12} + F_{13} + F_{23} \le 1 + 2 \sqrt{F_{12} F_{13} F_{23}}.
 \label{ineq}
 \end{equation} 
\end{lemma}

\begin{proof}
 The three dimensional matrix 
 \begin{equation*}
  G =
  \begin{pmatrix}
  1 &\<\varphi_1,\varphi_2\> &\<\varphi_1,\varphi_3\> \\
  \<\varphi_2,\varphi_1\> &1 &\<\varphi_2,\varphi_3\> \\
  \<\varphi_3,\varphi_1\> &\<\varphi_3,\varphi_2\> &1 
  \end{pmatrix}
 \end{equation*}
 is positive semi-definite. In particular, $\det(G) \ge 0$  or
 \begin{equation*}
  0 \le 1 + 2 \Re\g e\Bigl(\<\varphi_1,\varphi_2\> \<\varphi_2,\varphi_3\>
  \<\varphi_3,\varphi_1\> \Bigr) - \abs{\<\varphi_1,\varphi_2\>}^2 -
  \abs{\<\varphi_1,\varphi_3\>}^2- \abs{\<\varphi_2,\varphi_3\>}^2.
 \end{equation*}
Hence
\begin{align*}
 F_{12} + F_{13} + F_{23}
 &= \abs{\<\varphi_1,\varphi_2\>}^2 + \abs{\<\varphi_1,\varphi_3\>}^2 +
  \abs{\<\varphi_2,\varphi_3\>}^2 \\
 &\le 1 + 2 \Re\g e\Bigl(\<\varphi_1,\varphi_2\> \<\varphi_2,\varphi_3\>
  \<\varphi_3,\varphi_1\> \Bigr) \\
 &\le 1 + 2 \abs{\<\varphi_1,\varphi_2\>} \abs{\<\varphi_2,\varphi_3\>}
  \abs{\<\varphi_3,\varphi_1\>}  \\
 &= 1 + 2 \sqrt{F_{12} F_{13} F_{23}},
\end{align*} 
which proves~(\ref{ineq}).
\end{proof}

Let us denote by $\c C_3$ the compact convex subset of $\Rl^3$
\begin{equation*}
 \{\bi x = (x_1,x_2,x_3) \mid 0\le x_j\le1 \text{ for } j=1,2,3 \text{ and }
 x_1^2+x_2^2+x_3^3 \le 1 + 2 x_1 x_2 x_3 \}.
\end{equation*}
The convex boundary of $\c C_3$ is 
\begin{align*}
 \partial_{\r c}(\c C_3) = 
 &\{(0,0,0), (0,0,1), (0,1,0), (1,0,0), (1,1,1)\} \\
 &\cup \{\bi x \mid 0\le x_j<1 \text{ for } j=1,2,3 \text{ and } x_1^2+x_2^2+x_3^2
 = 1 + 2x_1x_2x_3\}.
\end{align*}

\begin{lemma}
\label{lem2}
 $\c C_3 = \{ (F_{12}^{\frac{1}{2}}, F_{13}^{\frac{1}{2}},
 F_{23}^{\frac{1}{2}})\}$ where the fidelities are computed for probability
 measures on a space with six points.
\end{lemma}

\begin{proof}
The extreme point $(0,0,0)$ can be realised with three degenerate measures
on a space of three points, $(1,0,0)$, $(0,1,0)$ and $(0,0,1)$ on one with
two points and $(1,1,1)$ with a single point.  Using convexity and the shape
of $\c C_3$, it is then sufficient to show that any extreme point of the
surface $\{\bi x \mid 0\le x_j<1 \text{ for } j=1,2,3 \text{ and } x_1^2+x_2^2+x_3^2
= 1 + 2x_1x_2x_3\}$ can be obtained by transition probabilities between
measures on a space with two points.

Suppose that $x_1\le x_2\le x_3$ and put $\lambda_1 := (1,0)$, $\lambda_2 :=
(x_1^2,1-x_1^2)$ and $\lambda_3 := (x_2^2,1-x_2^2)$. We then have that
\begin{align*}
 &F(\lambda_1 \,; \lambda_2) = x_1^2,\quad 
 F(\lambda_1 \,; \lambda_3) = x_2^2,\quad\text{and}\\
 &F(\lambda_2 \,; \lambda_3) = \bigl( x_1 x_2 + \sqrt{(1-x_1^2)(1-x_2^2)}
 \bigr)^2.
\end{align*}
Solving the equation $x_1^2+x_2^2+x_3^2 = 1 + 2x_1x_2x_3$ for $x_3$ and
taking into account that $0\le x_1\le x_2\le x_3\le 1$ we find that
$F(\lambda_2 \,; \lambda_3) = x_3^2$.
\end{proof}

Lemma~\ref{lem2} shows that any triple of transition probabilities that can
be obtained by transition probabilities between pure quantum states can also
be reached by classical measures. Our next aim is to show the rather
surprising fact that we still remain in the set $\c C_3$ when we consider 
Uhlmann fidelities between three arbitrary mixed quantum states. 

\begin{proposition}
\label{prop1}
 Let $\rho_1$, $\rho_2$ and $\rho_3$ be three density matrices on 
 a Hilbert space $\g H$ and let $F_{jk} := F(\rho_j \,; \rho_k)$, then
 \begin{equation*}
  F_{12} + F_{13} + F_{23} \le 1 + 2 \sqrt{F_{12} F_{13} F_{23}}.
 \end{equation*} 
\end{proposition}
   
\begin{proof}
We may assume that the $\rho_j$ have full support. The general case is 
obtained by continuity. Denoting $F(\rho_1 \,; \rho_2)^{\frac{1}{2}}$ by
$a$, we certainly have that $0<a\le 1$. Therefore
\begin{equation*}
 0 < F(\rho_1 \,; \rho_3) + F(\rho_2 \,; \rho_3) - 2 a \sqrt{F(\rho_1 \,;
 \rho_3)\, F(\rho_2 \,; \rho_3)}
\end{equation*} 
and we wish to obtain the upper bound $1-a^2$. Using the polar decomposition
and the assumption on the support of the $\rho_j$, there exist unitaries $U$
and $V$ such that
\begin{equation*}
 \rho_1^{\frac{1}{2}}\, \rho_3^{\frac{1}{2}} = U\, \abs{\rho_1^{\frac{1}{2}}\, 
 \rho_3^{\frac{1}{2}}}
 \qquad\text{and}\qquad
 \rho_2^{\frac{1}{2}}\, \rho_3^{\frac{1}{2}} = V\, \abs{\rho_2^{\frac{1}{2}}\, 
 \rho_3^{\frac{1}{2}}}.
\end{equation*} 
In these formulas $\abs{X}$ denotes the absolute value of an operator $X$,
i.e.\ $\abs{X} := (X^*X)^{\frac{1}{2}}$. We now express the fidelities as
follows
\begin{equation*}
 F^{\frac{1}{2}}(\rho_1 \,; \rho_3) = \tr \sqrt{\rho_3^{\frac{1}{2}}\,
 \rho_1\, \rho_3^{\frac{1}{2}}} = \tr \abs{\rho_1^{\frac{1}{2}}\, \rho_3^{\frac{1}{2}}}
 = \tr U^* \rho_1^{\frac{1}{2}}\, \rho_3^{\frac{1}{2}}.
\end{equation*}
Using the Hilbert-Schmidt scalar product between Hilbert-Schmidt operators,
$\< X,Y \>_{\r{HS}} := \tr X^*Y$, the fidelities become
\begin{equation*}
 F^{\frac{1}{2}}(\rho_1 \,; \rho_3) = \<f,h\>_{\r{HS}}
 \qquad\text{and}\qquad
 F^{\frac{1}{2}}(\rho_2 \,; \rho_3) = \<g,h\>_{\r{HS}}
\end{equation*} 
with 
\begin{equation*}
 f := \rho_1^{\frac{1}{2}}\,U,\quad
 g := \rho_2^{\frac{1}{2}}\,V\quad \text{and }
 h := \rho_3^{\frac{1}{2}}.
\end{equation*}
We can then verify the following properties
\begin{align*}
 &\<f,h\>_{\r{HS}} = \abs{\<f,h\>_{\r{HS}}},\ \<g,h\>_{\r{HS}} =
 \abs{\<g,h\>_{\r{HS}}}, \\
 &\norm{f}_{\r{HS}} = \norm{g}_{\r{HS}} = \norm{h}_{\r{HS}} = 1,\quad
 \text{and} \\
 &\abs{\<f,g\>} = \abs{\tr U^* \rho_1^{\frac{1}{2}} \rho_2^{\frac{1}{2}} V}
 \le \sup_{W\ \text{unitary}}\ \abs{\tr \rho_1^{\frac{1}{2}} \rho_2^{\frac{1}{2}}
 W} = F^{\frac{1}{2}}(\rho_1 \,; \rho_2) = a.   
\end{align*}
The statement of the proposition now follows from Lemma~\ref{lem3}.
\end{proof}

\begin{lemma}
\label{lem3}
Let $f$ and $g$ be normalised vectors in a Hilbert space $\g H$ and let $a$
be such that $\abs{\<f,g\>} \le a \le 1$, then
\begin{equation*}
 \sup_{h,\ \norm{h}\le1 } \Bigl( \abs{\<f,h\>}^2 + \abs{\<g,h\>}^2 - 2 a
 \abs{\<f,h\>} \abs{\<g,h\>} \Bigr) = (1-a)(1+\abs{\<f,g\>}) \le (1-a^2).
\end{equation*}
\end{lemma}

\begin{proof}
As the supremum is always non-negative, we may impose the additional restriction
$h\in\spa(\{f,g\})$. If $h$ does not belong to this subspace, decompose it
into $h_1\oplus h_2$ with $h_1\in\spa(\{f,g\})$. Next replace $h$ by
$h_1/\norm{h_1}$. Evaluating the functional with this new $h$ will return a
value at least as large as that with the original $h$. 

So let $h = \alpha f + \beta g$ with $\alpha,\beta\in\Cx$ such that
\begin{equation}
 \norm{h}^2 = \abs{\alpha}^2 + \abs{\beta}^2 + 2 \Re\g e(\overline\alpha
 \beta \<f,g\>) = 1.
\label{nor}
\end{equation}
Using this normalisation condition we compute
\begin{equation*}
 \abs{\<f,h\>}^2 
 = \abs{\alpha + \beta \<f,g\>}^2 
 = 1 - \abs{\beta}^2 \bigl( 1-\abs{\<f,g\>}^2\bigr)
\end{equation*}
and
\begin{equation*}
 \abs{\<g,h\>}^2 
 = \abs{\overline\alpha \<f,g\> + \overline\beta}^2 
 = 1 - \abs{\alpha}^2 \bigl( 1-\abs{\<f,g\>}^2\bigr).
\end{equation*}
Hence, the functional of $h$ we have to maximise does not depend on the
phase of $\overline\alpha \beta \<f,g\>$. The normalisation
condition~(\ref{nor}) can be satisfied if and only if
\begin{equation*}
 \Bigl| \abs{\alpha}^2 + \abs{\beta}^2 -1 \Bigr| \le 2 \abs{\alpha}
 \abs{\beta} \abs{\<f,g\>}.
\end{equation*} 
Putting $\lambda := \abs{\alpha}$, $\mu := \abs{\beta}$ and $t := \abs{\<f,g\>}$
we have to compute
\begin{equation*}
 I := \sup_{\lambda,\mu} \Bigl( 2 - (\lambda^2+\mu^2)(1-t^2) - 2a
\sqrt{1-\lambda^2(1-t^2)} \sqrt{1-\mu^2(1-t^2)} \Bigr)
\end{equation*} 
subject to the constraints
\begin{equation*}
 0\le\lambda,\quad 0\le\mu,\quad \text{and } \abs{\lambda^2 + \mu^2-1} \le
 2\lambda\mu t
\end{equation*}
with $t$ satisfying $0\le t\le a\le 1$. The supremum is attained choosing
$\lambda=\mu$, with $\lambda$ such that
\begin{equation*}
 \frac{1}{2(1+t)} \le \lambda^2 \le \frac{1}{2(1-t)}.
\end{equation*}
We obtain for $I$ the value
\begin{equation*}
 I = 2 (1-a) \Bigl(1 -\frac{1-t^2}{2(1+t)} \Bigr) = (1-a)(1+t) \le (1-a^2).
\end{equation*}
\end{proof}

\section{Phase, a three state invariant}
\label{s3}

For three pure states $\omega_j := P_{\varphi_j}$ with $\varphi_j\in\g H$ and
$\norm{\varphi_j}=1$, $j=1,2,3$ we introduce the complex number 
$\r e^{i\Phi(\omega_1;\omega_2;\omega_3)}$ by the relation
\begin{align*}
 \< \varphi_1,\varphi_2\> \< \varphi_2,\varphi_3\>\< \varphi_3,\varphi_1\> 
 &= \r e^{i\Phi(\omega_1; \omega_2; \omega_3)} \Bigl| \< \varphi_1,\varphi_2\> 
 \< \varphi_2,\varphi_3\>\< \varphi_3,\varphi_1\> \Bigr| \\
 &= \r e^{i\Phi(\omega_1; \omega_2; \omega_3)} F^{\frac{1}{2}}(\omega_1 \,; \omega_2) 
 F^{\frac{1}{2}}(\omega_2 \,; \omega_3) F^{\frac{1}{2}}(\omega_3 \,;
 \omega_1).
\end{align*}
This definition implicitly assumes that all fidelities are strictly
positive. It is easy to see that there is no continuous extension to the
case of zero fidelities. Obviously $\Phi$ is invariant under isometric
transformations and
\begin{equation*}
 \Phi(\omega_1;\omega_2;\omega_3) = \Phi(\omega_2;\omega_3;\omega_1)
 \qquad\text{and}\qquad
 \Phi(\omega_1;\omega_2;\omega_3) =
 -\Phi(\omega_2;\omega_1;\omega_3).
\end{equation*}
The relevance of other invariants than the fidelity was e.g.\ underlying the study
of the spectrum of the Gram matrix of a sequence of states as
in~\cite{decfanspin}. An other instance was considered in~\cite{mitjoz} where
compressibility was shown to depend monotonically on the elementary
symmetric invariants of the Gram matrix. Here we shall argue the need for
such an invariant to reconstruct sequences of pure states, up to unitary
invariance of course.

Let us first consider the example of a two level system. Let $\bi x$ be a
unit vector in $\Rl^3$ that defines through
the Bloch transformation a pure state on the $2\times2$ matrices. Let us fix
in the canonical basis of $\Cx^2$ a vector $\varphi_{\bi x}$ that generates 
the state. Using standard spherical coordinates $\bi x = (\cos\phi
\sin\theta, \sin\phi \sin\theta, \cos\theta)$ we choose 
\begin{equation*}
 \varphi_{\bi x} := \begin{pmatrix} \cos \frac{\theta}{2} \\ \r e^{i\phi}
 \sin \frac{\theta}{2} \end{pmatrix}, \quad \theta\in(0,\pi),\
 \phi\in(0,2\pi(.
\end{equation*}
Fix now three unit vectors $\bi x$, $\bi y$ and $\bi z$ in $\Rl^3$. 
A straightforward computation shows that 
\begin{equation*}
 \cos\bigl(\Phi(\rho_{\bi x};  \rho_{\bi y}; \rho_{\bi z})\bigr) 
 = \frac{ \cos^2( \frac{\theta_{\bi{xy}}}{2} ) +
 \cos^2( \frac{\theta_{\bi{yz}}}{2} ) + \cos^2( \frac{\theta_{\bi{zx}}}{2} ) -
 1}{2 \cos( \frac{\theta_{\bi{xy}}}{2}) \cos( \frac{\theta_{\bi{yz}}}{2} )
 \cos( \frac{\theta_{\bi{zx}}}{2}) }.
\end{equation*}
Here, $\theta_{\bi{xy}}$ denotes the angular distance between $\bi
x$ and $\bi y$. 
We need of course also $\sin\bigl(\Phi(\rho_{\bi x};  \rho_{\bi y}; \rho_{\bi
z})\bigr)$
in order to determine $\Phi$ uniquely. The angle $\Phi$ turns out to be zero
if and only if there is a permutation $(\alpha,\beta,\gamma)$ of
$(\bi{xy},\bi{yz},\bi{zx})$ such that $\theta_\alpha = \theta_\beta -
\theta_\gamma$, i.e. if $\bi x$, $\bi y$ and $\bi z$ are collinear.

Consider next the following rather trivial problem. The fidelity between two
degenerate classical probabilities is either one or zero according to
whether the points in which the measures live coincide or not. Suppose that
we are given for a sequence $\bi\delta = \{\delta_1, \delta_2, \ldots, \delta_n\}$
of Dirac measures the pairwise fidelities $F_{jk} := F(\delta_j \,; \delta_k)$.
We can obviously reconstruct the
sequence $\bi\delta$, up to renaming points and reshuffling. 
It suffices to determine the multiplicity of each
$\delta_j$ in the sequence by counting the number of $k$'s such that
$F_{jk}=1$. The analogous quantum problem is less trivial.

\begin{proposition}
For $n=1,2,\ldots$ there is generically, up to isometric transformations and reshuffling, 
a unique sequence $\bi\omega = (\omega_1, \omega_2, \ldots, \omega_n)$ of
pure states on a Hilbert space $\g H$ with given fidelities $F_{jk} :=
F(\omega_j \,; \omega_k)$ ($j<k$) and phases $\Phi_{jk\ell} :=
\Phi(\omega_j;
\omega_k; \omega_\ell)$ ($j<k<\ell$).
\end{proposition}      

\begin{proof}
Writing $\omega_j = P_{\varphi_j}$, we shall reconstruct the $\varphi_j$ in
a canonical way. More precisely
\begin{equation*}
\varphi_j = \sum_{\ell=1}^j c_{j\ell}\, e_\ell 
\end{equation*} 
where $\{e_1, e_2, \ldots\}$ is the standard basis in $\ell^2(\Nl_0)$, up
to multiplication of $e_\ell$ by a complex number $z_\ell$ of modulus one.
As $\varphi_j$ is only determined up to a phase and using the freedom to
choose the phases $z_\ell$ of the standard basis, we may assume that
$c_{j1}>0$ and $c_{jj}>0$. It remains to determine the $c_{jk}$ with
$1<k<j$. This can be done recursively. We start with $c_{11}=1$. For $j=2$
\begin{equation*}
c_{21} = F^{\frac{1}{2}}_{12}
\qquad\text{and}\qquad
c_{22} = \bigl( 1 - F_{12} \bigr)^{\frac{1}{2}}.
\end{equation*} 
For $j=3$, $c_{31} = F^{\frac{1}{2}}_{13}$. In order to determine the
complex coefficient $c_{32}$, we use the data $F_{13}$ and $\Phi_{123}$
\begin{equation*}
 \r e^{-i\Phi_{123}} \<\varphi_1,\varphi_2\> \<\varphi_2,\varphi_3\>
 \<\varphi_3,\varphi_1\> =
 F^{\frac{1}{2}}_{12}\, F^{\frac{1}{2}}_{13}\, F^{\frac{1}{2}}_{23}. 
\end{equation*}
This equation determines $\<\varphi_2,\varphi_3\>$ and from there on $c_{32}$
as
\begin{equation*}
 \<\varphi_2,\varphi_3\> = c_{21} c_{31} + c_{22} c_{32}.
\end{equation*} 
Given the coefficients $c_{k\ell}$ with $1\le k<j$ and $1\le \ell\le k$, 
we determine uniquely the $c_{j\ell}$ with $1\le\ell\le j$ by the equations
\begin{align*}
&\bullet\qquad c_{j1} = F^{\frac{1}{2}}_{1j} \\
&\bullet\qquad \r e^{-i\Phi_{12j}} \<\varphi_1,\varphi_2\>
\<\varphi_2,\varphi_j\> \<\varphi_j,\varphi_1\> = F^{\frac{1}{2}}_{12}\,
F^{\frac{1}{2}}_{1j}\, F^{\frac{1}{2}}_{2j} \\ 
&\bullet\qquad \r e^{-i\Phi_{13j}} \<\varphi_1,\varphi_3\>
\<\varphi_3,\varphi_j\> \<\varphi_j,\varphi_1\> = F^{\frac{1}{2}}_{13}\,
F^{\frac{1}{2}}_{1j}\, F^{\frac{1}{2}}_{3j} \\
&\phantom{\bullet\qquad\ }\vdots \\
&\bullet\qquad c_{jj} = \Bigl( 1 - \abs{c_{j1}}^2 - \cdots- \abs{c_{jj-1}}^2\Bigr)^{\frac{1}{2}} 
\end{align*}
\end{proof}

It is already clear from the three state case that only specifying the
fidelities of pairs in a sequence of pure states is insufficient to
reconstruct the sequence. A simple parameter count shows that we need
$(n-1)^2$ real parameters for a sequence of length $n$. The fidelities
provide $n(n-1)/2$ parameters and we therefore need $(n-1)(n-2)/2$ more. The
whole set of phases provides $n(n-1)(n-2)/6$ which is far too much for
large $n$. The argument of the proof is however based on the specification
of phases $\Phi_{1kj}$ with $1<k<j$ which is the minimal number needed.
 
To conclude, we propose an extension of the phase to the mixed state case
based on the purification of states. To keep things as simple as possible we
shall only deal with density matrices $\rho$ on $\g H$ with non-degenerate
spectrum and trivial kernel.
The general case can be dealt with taking the support of the density matrix
into account. We briefly remind here the purification construction, also
known as GNS~construction. A complex conjugation on $\g H$ is a complex
antilinear transformation $\varphi \mapsto \overline\varphi$ such that
\begin{equation*}
 \overline{\overline{\varphi}} = \varphi
 \qquad\text{and}\qquad
 \<\overline\varphi,\overline\psi\> = \<\psi,\varphi\>.
\end{equation*} 
To any operator $A$ on $\g H$ we may associate a conjugate operator
$\overline A$ by the relation $\overline A \varphi := 
\overline{A\overline\varphi}$. It is straightforward to check that
$\overline A$ is a complex linear operator on $\g H$ and that
\begin{equation*}
 \overline{\alpha A+B} = \overline\alpha\, \overline A + \overline B, \qquad
 \overline{AB} = \overline A\overline B, \qquad\text{and}\qquad
\overline{A^*} = \bigl( \overline A \bigr)^*. 
\end{equation*} 

Let $\{f_1, f_2, \ldots\}$ be an orthonormal basis of $\g H$ diagonalising
$\rho$: $\rho f_j = r_j\, f_j$. As standard purification of $\rho$, we
consider the vector
\begin{equation}
 \Omega_\rho := \sum_j r_j^{\frac{1}{2}}\, f_j \otimes \overline{f_j}
\qquad\text{in } \g H\otimes\g H.
\label{schmidt}
\end{equation}
Then $\rho$ is the marginal of the pure state $X \mapsto \<\Omega_\rho,
X\,\Omega_\rho\>$. The expansion of $\Omega_\rho$ in~(\ref{schmidt}) is the
Schmidt decomposition of $\Omega_\rho$. There are many pure states in the
tensor system that have $\rho$ as a marginal. They are all of the form
$\idty\otimes\overline U\, \Omega_\rho$ with $U$ a unitary operator on $\g
H$. We shall use this freedom of choice to extend the notion of phase from
pure states to mixed ones, namely $\Phi(\rho_1 ; \rho_2 ; \rho_3)$ 
is such that
\begin{equation}
 \Bigl| \r e^{i\Phi(\rho_1 ; \rho_2 ; \rho_3)} -1 \Bigr| =
 \inf_{U_1,U_2,U_3} \Bigl| \r e^{i\Phi(\idty\otimes\overline{U_1}\,\Omega_1 
 \,, \idty\otimes\overline{U_2}\,\Omega_2 \,,
\idty\otimes\overline{U_3}\,\Omega_3)} -1 \Bigr|.
\label{def}
\end{equation} 
Here the infimum is taken over all unitaries on $\g H$. 

We can rephrase the variational problem~(\ref{def}). Let $\{g_1, g_2, \ldots \}$
be the orthogonal basis of eigenvectors of a density matrix $\sigma$ with
corresponding eigenvalues $s_j$ and let $V$ be another unitary on $\g H$, then
\begin{align*}
 \<\idty\otimes\overline U\, \Omega_\rho, \idty\otimes\overline V\,
 \Omega_\sigma\>
 &= \sum_{jk} r_j^{\frac{1}{2}} s_k^{\frac{1}{2}} \<f_j\otimes\overline{U\,f_j},
 g_k\otimes\overline{V\,g_k}\> \\
 &= \sum_{jk} r_j^{\frac{1}{2}} s_k^{\frac{1}{2}} \<f_j, g_k\> \<V\,g_k,U\,f_j\> \\
 &= \tr \rho^{\frac{1}{2}} \sigma^{\frac{1}{2}} V^*U.
\end{align*}
A study of the phase for mixed states will be the subject of a forthcoming
paper.

\noindent
\textbf{Acknowledgements}
It is a pleasure to acknowledge stimulating and interesting discussions with
R.~Alicki, M.~Horodecki, R.~Horodecki and P.~Spincemaille. This work was
partially suported by F.W.O., Vlaanderen grant G.0109.01.


\begin{thebibliography}{99}
\bibitem{ararag}
H.~Araki and G.A.~Raggio:
A remark on transition probability,
\emph{Lett.\ Math.\ Phys.\ }\textbf{6}, 237--240 (1982) 
\bibitem{decfanspin}
 M.~De~Cock, M.~Fannes and P.~Spincemaille: 
 On quantum dynamics and statistics of vectors,  
 \emph{J.\ Phys.\ A} \textbf{32}, 6547--6571 (1999)
\bibitem{fanspin}
 M.~Fannes and P.~Spincemaille: 
 Multiple return times in the quantum baker map, 
 \emph{Phys.\ Lett.\ A } \textbf{294}, 74--78 (2002)
\bibitem{fanspin2}
 M.~Fannes and P.~Spincemaille: 
 The mutual affinity of random measures, 
 \emph{Periodica Mathematica Hungarica}, \textbf{47}, 51--71 (2003)
\bibitem{mitjoz}
G.~Mitchison and R.~Jozsa:
Towards a geometrical interpretation of quantum information
compression
e-Print: quant-ph/0309177 
\bibitem{petmos}
D.~Petz and M.~Mosonyi:
Stationary quantum source coding,
\emph{J.\ Math.\ Phys.\ }\textbf{42}, 4857--4864 (2001)
\bibitem{sch}
B.~Schumacher:
Quantum coding,
\emph{Phys.\ Rev.\ A} \textbf{51}, 2738--2747 (1995)
\bibitem{uhl}
 A.~Uhlmann:
 The transition probability in the state space of a $\ast$-algebra,
\emph{Rep.\ Math.\ Phys.\ }\textbf{9}, 273--279 (1976)

\end{thebibliography}
\end{document}